\documentclass[smallextended]{svjour3}

\usepackage{url}
\usepackage{graphicx}
\usepackage{subfigure}
\usepackage{color}
\usepackage{tablefootnote}
\usepackage{footnote}
\usepackage{multirow}

\begin{document}

\title{The H-index Paradox: Your Coauthors Have a Higher H-index than You Do\\
{\tiny Accepted for publication in Scientometrics}}

\author{Fabr\'icio Benevenuto \and \\
        Alberto H. F. Laender  \and \\
        Bruno L. Alves
}

\institute{Computer Science Department \\
Federal University of Minas Gerais \\
\email{\{fabricio, laender, bruno.leite\}@dcc.ufmg.br}\\
Tel.: +5531-3409-5860\\
Fax: +5531-3409-5858\\
}

\maketitle
\begin{abstract}
\begin{quote}
One interesting phenomenon that emerges from the typical structure of social networks is the {\it friendship paradox}. It states that your friends have on average more friends than you do. Recent efforts have explored variations of it, with numerous implications for the dynamics of social networks. However, the friendship paradox and its variations consider only the topological structure of the networks and neglect many other characteristics that are correlated with node degree. In this article, we take the case of scientific collaborations to investigate whether a similar paradox also arises in terms of a researcher's scientific productivity as measured by her H-index. The H-index is a widely used metric in academia to capture both the quality and the quantity of a researcher's scientific output. It is likely that a researcher may use her coauthors' H-indexes as a way to infer whether her own H-index is adequate in her research area. Nevertheless, in this article, we show that the average H-index of a researcher's coauthors is usually higher than her own H-index. We present empirical evidence of this paradox and discuss some of its potential consequences.
\end{quote}
\end{abstract}

\section{Introduction}

One interesting phenomenon that emerges from the typical structure of social networks is the \textit{friendship paradox}~\cite{quatro}. It states that, on average, your friends have more friends than you do. This paradox basically exists because of the discrepancy on node degree values in typical social networks~\cite{barabasi1999emergence}, in which individuals with a high number of friends are over-represented when averaging over them~\cite{hodas2014icwsm}. As a consequence, the friendship paradox can dramatically skew an individual's local observation, making such an observation appear far more common than it is in reality~\cite{centola2010spread,salganik2006experimental}.

In this context, identifying variations of this paradox in different ecosystems has been the topic of some important recent research efforts~\cite{eom2014generalized,tres,lerman2015majority}. For instance, two new paradoxes have been verified on Twitter~\cite{tres}: (1) the \textit{virality paradox} that states that your friends receive more viral content than you do, and (2) the \textit{activity paradox} that states that your friends post more frequently than you do. More recently, the friendship paradox was generalized to any complex network~\cite{eom2014generalized} and its origins are highly correlated with the skewed distribution of node degree (i.e., the number of network friends)~\cite{hodas2014icwsm}. In a nutshell, these efforts suggest that any attribute that is highly correlated with node degree is likely to produce this kind of paradox~\cite{eom2014generalized}. Thus, in this article we take the case of scientific collaborations to investigate whether a similar paradox also arises in terms of a researcher's scientific productivity as measured by her H-index.


The H-index~\cite{Hirsch:2005} is a metric originally proposed to measure a researcher's scientific output. Its calculation is quite simple as it is based on the researcher's set of most cited publications and the number of citations they have received.  More specifically, a researcher has an H-index $h$ if she has at least $h$ publications that have received at least $h$ citations. Thus, if a researcher has at least 10 publications with at least 10 citations, her H-index is 10.

Like any metric that attempts to summarize a complex and subjective evaluation in a single number, the H-index has its limitations, including being biased towards the researchers' scientific lifetime, not accounting for the number of coauthors in the publications and ignoring the distinct citation patterns across different areas~\cite{bornmann2005does}. Nevertheless, the H-index became popular as it provides a notion of both quality and quantity of a researcher's scientific output in a simple and easy-to-compute metric. As a consequence, researchers are often tempted to evaluate themselves based on the H-index. Systems like Google Scholar\footnote{\url{http://scholar.google.com/intl/en/scholar/citations.html}} and ArnetMiner\footnote{\url{http://arnetminer.org}} help researchers track their publication impact and coauthors, as well as to maintain their profiles, where the H-index is clearly stamped. Thus, it is natural to assume that researchers may use their coauthors' H-indexes as a way to estimate whether their own H-index is adequate in their respective research areas or within a department or university.

Despite recent efforts to generalize the friendship paradox~\cite{eom2014generalized}, it is still unclear whether a similar paradox actually happens when we consider the H-index in a coauthorship network. However, we have been able to show that the average H-index of a researcher's coauthors is usually higher than her own H-index.


Next, we briefly discuss how we have estimated the H-index for researchers from distinct Computer Science research communities, and then provide empirical results that corroborates the existence of the {\it H-index paradox}.

\section{Estimating H-index}

In order to provide evidence of the H-index paradox, we need to be able to (1) identify the coauthors of a large set of researchers and (2) estimate the H-index of these researchers as well as of their respective coauthors.

We focus on constructing the coauthorship network of Computer Science researchers from different areas. To do that, we gathered data from DBLP\footnote{\url{http://www.informatik.uni-trier.de/~ley/db/}}, as it offers its entire database in XML format for download. We gathered this data for those researchers who published in the flagship conferences of 10 major ACM SIGs (Special Interest Groups)\footnote{http://www.acm.org/sigs}: SIGCHI, SIGCOMM, SIGCSE, SIGDOC, SIGGRAPH, SIGIR, SIGKDD,  SIGMETRICS, SIGMOD and SIGPLAN.

There are several tools that measure the H-index of researchers, of which Google Scholar is today the most prominent one. However, in order to have a profile in this system, a researcher needs to sign up and explicitly create it. In a preliminary collection of part of the profiles of the DBLP authors, we found that less than 30\% of these authors had a profile at Google Scholar~\cite{um}. Thus, this strategy would largely reduce our dataset.

To overcome this limitation, we used data from the SHINE (Simple HINdex Estimation) project\footnote{\url{http://shine.icomp.ufam.edu.br}} to estimate the researchers' H-index. SHINE provides a website that shows the H-index of almost 1,800 Computer Science conferences. It was created based on a large scale crawl of Google Scholar. Its strategy consisted of searching for the title of all papers published in such conferences, thus effectively estimating their H-index based on the citations computed by Google Scholar. Although SHINE only allows searching for the H-index of conferences, their developers kindly allowed us to use its dataset to infer the H-index of researchers based on the citations received by their conference papers.

\begin{figure}[!htb]
\centering
 \includegraphics[width=.99\textwidth]{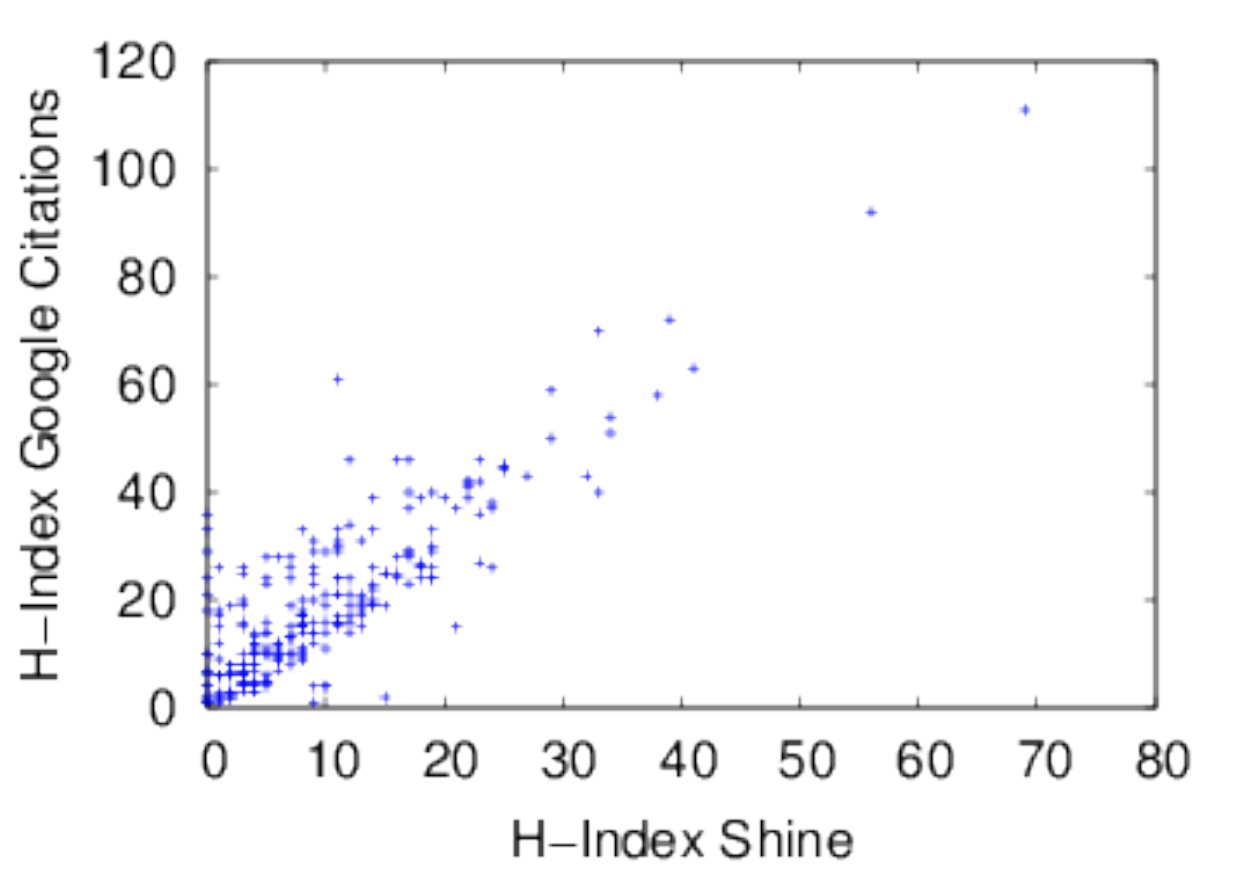}
\caption{Correlation between the inferred H-index and Google Citations one.}
\label{fig:hindex_scatter_plot}
\end{figure}

However, besides covering only conferences, SHINE does not track all existing conferences in Computer Science, which might cause the researchers' H-index to be underestimated when computed with this data. To investigate this issue, we compared the H-index of researchers with a profile on Google Scholar with their estimated H-index based on the SHINE data. For this, we randomly selected 10 researchers for each of the ACM SIG's flagship conferences and extracted their H-indexes from their respective Google Scholar profiles. In comparison with the H-index we estimated from SHINE, the Google Scholar values are, on average, 50\% higher. Figure~\ref{fig:hindex_scatter_plot} shows the scatterplot for the two H-index measures. We can note, however, that although the SHINE H-index is lower, the two measures are highly correlated. The Pearson's correlation coefficient is 0.85, indicating that the H-index estimations are proportional in both systems.

\begin{table*}[t]
\centering
\caption{The DBLP data of the 10 ACM SIG flagship conferences.}
\label{tab:sigs_conference_period}
{\small
\begin{tabular}{|l|l|c|c|c|c|c|} \hline
SIG & Conference & Period & H-Index & Authors & Publications & Editions \\ \hline
SIGDOC & SIGDOC & 1989-2010 & 23 & 1071 & 810 & 22 \\ \hline
SIGCHI & CHI & 1994-2012 & 144 & 5095 & 2819 & 19 \\ \hline
SIGIR & SIGIR & 1978-2011 & 116 & 3624 & 2687 & 34 \\ \hline
SIGKDD & KDD & 1995-2011 & 124 & 3078 & 1699 & 17 \\ \hline
SIGCOMM & SIGCOMM & 1988-2011 & 140 & 1593 & 796 & 24 \\ \hline
SIGCSE & SIGCSE & 1986-2012 & 51 & 3923 & 2801 & 27 \\ \hline
SIGGRAPH & SIGGRAPH & 1985-2003 & 119 & 1920 & 1108 & 19 \\ \hline
SIGMETRICS & SIGMETRICS & 1981-2011 & 71 & 2083 & 1174 & 31  \\ \hline
SIGPLAN & POPL & 1975-2012 & 85 & 1527 & 1217 & 38  \\ \hline
SIGMOD & SIGMOD & 1975-2012 & 147 & 4202 & 2669 & 38  \\ \hline
\end{tabular}
}
\end{table*}

Table~\ref{tab:sigs_conference_period} summarizes the collected data, including the SIG, the conference acronym, the period considered (some conferences had their period reduced to avoid a hiatus in the data), the conference's SHINE H-index and the total number of authors, publications and editions. This dataset is useful to our purposes, since it allows us to investigate the H-index paradox on real Computer Science communities, in which researchers might tend to compare themselves with their peers.

\section{Comparing the H-index of a Researcher with her Coauthors'}

\begin{figure*}[t]
  \centering
  \includegraphics[width=.99\textwidth]{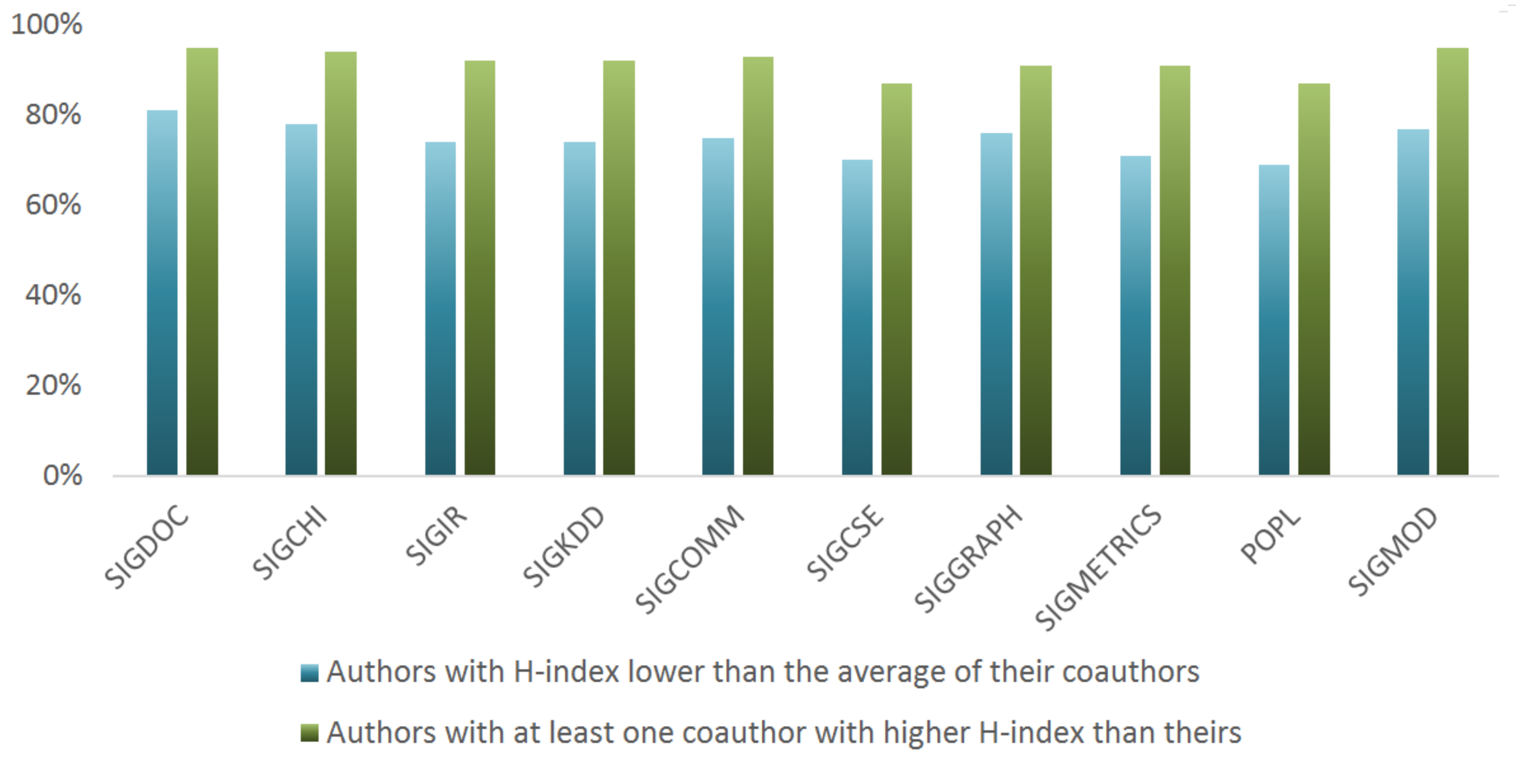}
  \caption{Comparison results of a researcher H-index with her coauthors.}
  \label{fig:comp}
\end{figure*}

Having estimated the H-index of each researcher, we can compare it with her coauthors'. Figure~\ref{fig:comp} shows the fraction of authors with an H-index that is lower than the average of their coauthors for the 10 conferences we have considered. We note that even focusing on authors that have published in flagship conferences of ACM SIGs, the fraction of authors that are below average is quite high for all research communities analyzed, varying from 69\% (POPL) to 81\% (SIGDOC). When we look at the percentage of authors with at least one coauthor with a higher H-index than theirs, the numbers are higher than 90\% for most of the conferences.

\begin{figure}[!htpb]
  \centering
  \includegraphics[width=.88\textwidth]{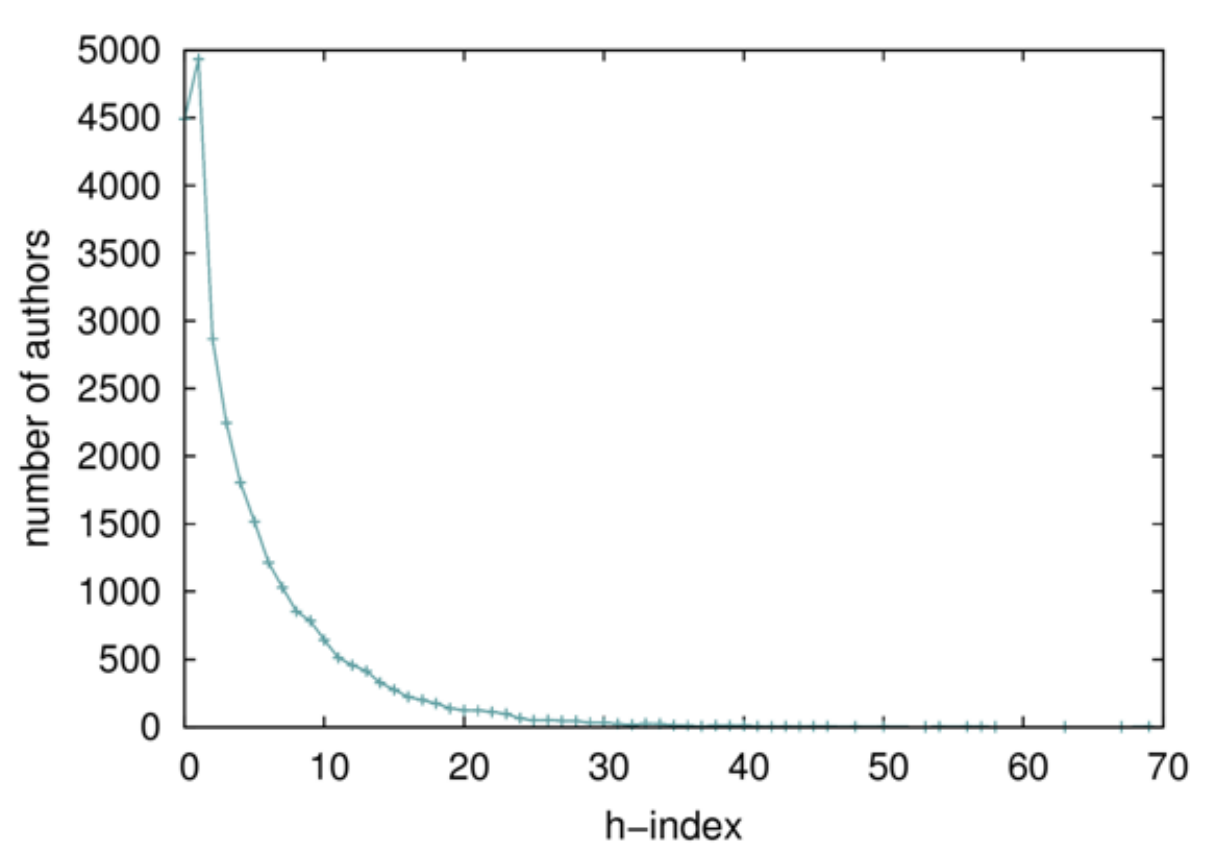}
  \caption{Distribution of authors according to their H-indexes.}
  \label{fig:distrib}
\end{figure}

These results confirm the H-index paradox since one's coauthors in a research community have, on average, a higher H-index than hers. The reasons behind the H-index paradox might be explained by the high correlation between node degree and H-index in a research community. Usually, high degree nodes tend to be senior researchers that not only advise a large number of students but also establish more collaborations, often with different groups along their career~\cite{um}. To further investigate this issue, Figure~\ref{fig:distrib} shows the distribution of the number of authors as a function of the H-index. It clearly resembles a long tail distribution, thus suggesting that some authors disproportionally contribute to the average H-index. This disproportion on the average H-index might be even sharper with the typical structural properties of coauthorship networks, which are similar to many social networks \cite{seis,oito}, i.e., they have a long tail degree distribution, in which highly connected authors create bridges across multiple highly connected components, leading to the properties of high clustering coefficient and short diameter. Finally, we measured the Pearson's coefficient correlation between a researcher's H-index and her degree. Such correlation is 0.36 a value that, although not very high, is positive, thus suggesting that a small number of researchers simultaneously have a high H-index and a large number of connections in the network.

\section{Conclusions}

In this article we have analyzed a variation of the well-known friendship paradox. By analyzing the average H-index of a researcher's coauthors for different Computer Science research communities, we show that the H-index paradox arises because the H-index is positively correlated with node degree. One of the implications of the friendship paradox is the fact that it leads to systematic biases in our perceptions. Thus, similarly, the H-index paradox induces researchers to feel that they rank below average in comparison with their coauthors.
This phenomenon is an instantiation of a sensation that occurs in different scenarios and is popularly captured by an expression that is common to many languages and cultures: \textit{the grass is always greener on the other side of the fence}~\cite{giansante2007grass}.


\section{Acknowledgments}
This research was partially funded by InWeb - The Brazilian National Institute of Science and Technology for the Web (MCT/CNPq/FAPEMIG grant 573871/2008-6), and by the authors' individual grants from CAPES, CNPq and FAPEMIG.

\bibliographystyle{spmpsci}

\end{document}